\def\year{2019}\relax
\documentclass[letterpaper]{article} %DO NOT CHANGE THIS
\usepackage{aaai19}  %Required
\usepackage{times}  %Required
\usepackage{helvet}  %Required
\usepackage{courier}  %Required
\usepackage{url}  %Required
\usepackage{graphicx}  %Required
\usepackage{thmtools, thm-restate}
\usepackage{subcaption}
\usepackage{bbm}

\usepackage{enumitem}
\usepackage{xcolor}
\usepackage{amsmath,amsfonts,amssymb, amsthm}
\usepackage{mathtools}

\makeatletter
\newcommand*\dashline{\rotatebox[origin=c]{90}{$\dabar@\dabar@\dabar@$}}
\makeatother

\newtheorem{definition}{Definition}
\newtheorem{proposition}{Proposition}
\newtheorem{lemma}{Lemma}
\newtheorem{theorem}{Theorem}
\newtheorem{corollary}{Corollary}

\definecolor{light-gray}{gray}{0.83}

\begin{document}

\title{Deep Bayesian Trust : A Dominant and Fair Incentive Mechanism for Crowd}

\author{Naman Goel and Boi Faltings\\
	Artificial Intelligence Laboratory\\
	\'Ecole Polytechnique F\'ed\'erale de Lausanne\\
	Lausanne, Switzerland, 1015\\
	\{naman.goel, boi.faltings\}@epfl.ch
}

\maketitle
\begin{abstract}
An important class of game-theoretic incentive mechanisms for eliciting effort from a crowd are the peer based mechanisms, in which workers are paid by matching their answers with one another. The other classic mechanism is to have the workers solve some {\em gold} standard tasks and pay them according to their accuracy on gold tasks. This mechanism ensures stronger incentive compatibility than the peer based mechanisms but assigning gold tasks to all workers becomes inefficient at large scale. We propose a novel mechanism that assigns gold tasks to only a few workers and exploits transitivity to derive accuracy of the rest of the workers from their peers' accuracy. We show that the resulting mechanism ensures a dominant notion of incentive compatibility and fairness.
\end{abstract}

\section{Introduction}\label{sec:intro}
Crowdsourcing is a popular method for collecting data but the collected data is often noisy and of low quality. The quality gets significantly degraded if solving the tasks require some costly effort. The problem can be addressed by rewarding the workers with a performance based bonus. One way to reward the workers is to use the peer based mechanisms~\cite{dasgupta2013crowdsourced,radanovic2016incentives}. In these mechanisms, the reward of a worker depends on her own answers and of other workers. The mechanisms admit honesty as an equilibrium strategy (i.e. if other workers do high quality work, the best response for any worker is also to do the same). But the mechanisms also admit other dishonest equilibria. The other way to reward the workers is to use gold standard tasks~\cite{oleson2011programmatic}. A common technique is to randomly mix some gold tasks in the batch of tasks solved by \textit{every} worker. Workers are then rewarded based on their performance on the gold tasks. This incentivizes the workers for performing high quality work as a dominant strategy (i.e. regardless of other workers' answers). However, since every worker solves gold tasks and the requester already knows the correct answers of the gold tasks, it leads to a waste of the useful task budget of the requester. Moreover, to reduce the variance in the rewards, one may require a sufficient number of gold tasks to be solved by every worker. Also, this technique only works if the gold tasks are not identified but as shown by \cite{checco2018all}, the gold tasks can be easily leaked because of their repeated use for all workers. Another technique is to assign gold tasks to a worker only with a small (constant) probability and give a fixed reward (independent of the quality of work) otherwise. While it solves many of the problems with the first technique, the scalability of the constant probability technique remains limited because the number of workers who need to be assigned gold tasks, grows as the total number of workers grows. Otherwise, this technique assumes that the rewards of the workers solving gold tasks can be made arbitrarily large to compensate for the smaller probability\footnote{This can be understood using the example of penalty mechanism in public transport systems. If tickets are checked very rarely, then the penalty for being found without ticket has to be made very high to discourage rational people from traveling without tickets.}. In many practical settings, this is undesired or not possible.

In this paper, we introduce a scalable incentive mechanism (the \textbf{\textit{Deep Bayesian Trust} Mechanism}) which guarantees strong incentive compatibility while assigning gold tasks only to a small (constant) number of workers and rewarding the rest using the answers provided by peer workers. Since we ensure that gold tasks are assigned to a constant \textit{number} of workers and not just with a constant \textit{probability}, our mechanism is also suitable for very large scale settings. Moreover, this mechanism doesn't suffer from the problem of arbitrarily large payments discussed earlier. When there is a non-zero cost of effort for solving the tasks, the mechanism still ensures the desired theoretical properties if the payments are scaled appropriately only to cover the cost of effort. This scaling constant doesn't depend on the probability of assigning gold tasks. The mechanism is based on the observation that in large scale crowdsourcing settings, every worker reports answers to many similar tasks and hence the joint distribution of the answers of any two workers can be used to infer the accuracy of one worker given the accuracy of the other worker. It starts by rewarding a small set of workers based on gold tasks and then uses the answers provided by the workers on non-gold tasks as \textit{contributed} gold tasks to reward more workers. It continues this \textit{deep} chain of trust to an arbitrary depth, until all the tasks have been solved by the required number of workers.

As fairness of algorithms affecting humans is becoming a critical issue, it is important to justify the fairness of algorithms that determine payments of the workers. We, for the first time, address the issue of fairness of crowdsourcing incentive mechanisms in a principled manner and show that our mechanism ensures fair rewards to the workers. The summary of our main contributions is as follows:

\begin{itemize}
	\item We propose a dominant uniform strategy incentive compatible (DUSIC) mechanism, called the \textbf{\textit{Deep Bayesian Trust} Mechanism}, which rewards a constant number of workers with gold tasks and the rest using peer answers.
	\item On one hand, our mechanism addresses the issues with existing gold tasks based mechanisms and on the other hand, it also shows how the limitations of purely peer based incentive mechanisms can be overcome in some cases by assigning gold tasks to a few workers. Thus, it is also of interest for the peer-prediction community.
	\item We define a notion of fairness of rewards in crowdsourcing and show that our mechanism ensures fairness.
	\item Through numerical experiments, we show the robustness of our mechanism under various reporting strategies of the workers. In a preliminary study conducted on Amazon Mechanical Turk, we observe that the mechanism helps in eliciting effort and improving the quality of responses.
\end{itemize}

\noindent The supplementary material for this paper is available on authors' website.

\section{Related Work}\label{sec:related}
The research on crowdsourcing incentive mechanisms is mainly divided into two categories. The first category of work assumes that some spot checking option (for example, gold standard tasks) is available. The constant probability mechanism, discussed in Section~\ref{sec:intro}, is analyzed formally by~\cite{gao-new}. This mechanism randomly selects a few workers and spot checks only those workers with an oracle. The rest of the workers are given a constant amount of reward (independent of the quality of their work). The scalability of this mechanism is limited because the number of workers who need to be spot checked, grows as the total number of workers grows. Otherwise, in order to compensate for the smaller probability of spot checking, the mechanism allows the payments of the spot checked workers to be arbitrarily large.

The second category of work assumes no gold tasks to be available and uses only peer answers. Such mechanisms are called the peer-consistency (or peer-prediction) mechanisms. The early mechanisms in this category were either not detail-free (required knowledge about the beliefs of the workers)~\cite{peer-prediction} or not minimal (required workers to also submit some additional information other than their answers on the tasks)~\cite{prelec2004bayesian,radanovic2013robust,witkowski2012robust}. On the other hand, a simple output-agreement mechanism~\cite{waggoner2014output} works only under strong assumptions on the correlation structure of workers' observations. A seminal work in the category of minimal, detail-free mechanisms for crowdsourcing is~\cite{dasgupta2013crowdsourced}, which ensures that truth-telling is a focal equilibrium in binary answer spaces. The Correlated Agreement mechanism~\cite{Shnayder2016} generalizes the mechanism of~\cite{dasgupta2013crowdsourced} to non-binary answer spaces with moderate assumptions on the correlation structure of workers' observations. Both these mechanisms require that workers solve multiple tasks. The Logarithmic Peer Truth Serum~\cite{goran-aamas15}, which is based on an information theoretic principle, requires no such assumptions and ensures strong-truthfulness in non-binary answer spaces. The guarantees of the mechanism are ensured in the limit (when every task is solved by an infinite number of workers). The Peer Truth Serum (PTSC) of~\cite{radanovic2016incentives} doesn't require even this assumption for the theoretical guarantees and works with a bounded number of tasks overall. In theory, these peer-consistency mechanisms offer comparatively weaker incentive compatibility than the gold tasks based mechanisms. They make truth-telling an equilibrium strategy for the workers but also admit some non-truthful equilibria. While the \textit{no-effort} or the \textit{heuristic equilibria} exist in these mechanisms, the equilibria are not attractive since they pay zero reward to the workers. The mechanisms also admit the \textit{permutation equilibria}, which give the same payoff as the truthful equilibrium. \cite{liu2018surrogate} avoid this issue in binary answer space by using ground truth of the answer statistics. As shown by~\cite{gao-new}, it is possible to eliminate the undesired equilibria in the peer based mechanisms if the center can employ a limited amount of spot checking. When spot checking is not possible, it is enough that there exist a small fraction of honest workers. Either of these options work if the rewards are scaled appropriately to compensate for a low probability of spot checking and a low fraction of honest agents respectively.

Finally, the mechanism of~\cite{de2016incentives} combines ideas from the two categories of work. It arranges the workers in a hierarchy. A constant number of workers in the top level of the hierarchy are evaluated by an oracle. The workers below that level are evaluated by the workers (peers) in one level above them. The mechanism solves the scalability issue of the gold tasks based mechanisms. Though it offers comparatively weaker incentive compatibility (unique Nash equilibrium) as compared to the gold tasks based mechanisms but eliminates all the undesired equilibria that exist in peer based mechanisms. However, it requires that workers are informed of their level in the hierarchy. The mechanism is also ex-ante unfair towards the workers in the sense that workers in the top level of hierarchy are evaluated more correctly than the workers in the lower levels. Similar to~\cite{de2016incentives}, our work is also at the intersection of the two categories of works. However, our mechanism doesn't suffer from the issues (level information requirement and unfairness) that their mechanism has and also guarantees stronger incentive compatibility.

\section{Model}
We consider large scale crowdsourcing settings in which workers provide answers of many micro-tasks requiring human intelligence. The tasks have a discrete answer space $\{0, 1, ..., K-1\}$ of size $K$. We will use $[K]$ to denote this space. For any task, our model has 3 random variables. The first is the unknown \textbf{ground truth} $G$ answer for the task. The second is the worker $i$'s \textbf{observed answer} $X_i$ that she obtains on solving the task. $X_i$ is worker's private information. The third is the worker's \textbf{reported answer} $Y_i$ that she actually reveals as her answer for the task. We use $g, x_i, y_i \in [K]$ to denote realizations of these random variables and will drop the subscript $i$, when the context is clear.

\begin{definition}[Effort Strategy]
	The effort strategy of a worker $i$ is a binary variable $e_i$. If the worker invests effort in solving a task,  $e_i$ is $1$ and is $0$ otherwise.
\end{definition}

The effort strategy captures the standard binary effort model of the incentive mechanisms literature~\cite{dasgupta2013crowdsourced}. Whenever $e_i = 1$, the worker incurs a strictly positive finite cost.

\begin{definition}[Reporting Strategy]
	When $e_i = 1$, the reporting strategy $S_i$ of a worker $i$ is a $K \times K$ right stochastic matrix, where $S_i[x,y]$ ($\forall x,y \in [K]$) is the probability of her reported answer on a task being $y$ given that her observed answer is $x$. When $e_i = 0$, the reporting strategy $\vec{S_i}$ of a worker is a $K$ dimensional probabilistic vector, where $\vec{S_i}[y]$ ($\forall y \in [K]$) is the probability of her reported answer on a task being $y$.
\end{definition}

The effort and the reporting strategy together model possible strategies that a worker may play in obtaining and reporting her answer and is a standard model in the literature~\cite{Shnayder2016}. Two common strategies are truthful and heuristic.

\begin{definition}[Truthful Strategy]
	A worker $i$'s strategy is called truthful if $e_i = 1$ and $S_i$ is an identity matrix.
\end{definition}

In a truthful strategy, a worker solves a task and reports her answer as obtained.

\begin{definition}[Heuristic Strategy]
	A worker $i$'s strategy is called heuristic either if $e_i = 0$ or if $e_i = 1$ and all rows of $S_i$ are identical.
\end{definition}

In a heuristic strategy, a worker either doesn't solve the tasks ($e_i = 0$) or solves the tasks ($e_i = 1$) but reports independently of the obtained answer. Note that a common colluding heuristic strategy, in which workers collude using a ``default" answer, is included in the model. For example, in binary case, when $e_i = 0$, a probabilistic vector $\vec{S_i} = [1,0]$ means that worker always answers $0$. Similarly, when $e_i = 1$, a matrix $S_i$ with both rows equal to $[1,0]$ means the same. It is also easy to see that the model also includes mixed strategies since mixed strategies can be written as convex combination of the pure strategies.

\begin{definition}[Proficiency Matrix]
	The proficiency matrix $A_i$ for a worker $i$ is a $K \times K$ right stochastic matrix, where $A_i[g,x]$ ($\forall g,x \in [K]$) is the probability of her obtained answer on a task being $x$ given that the ground truth is $g$.
\end{definition}

This definition is due to~\cite{dawid1979maximum}, which is a widely accepted model in crowdsourcing literature. The proficiency matrix models the ability of a worker to obtain correct answers, when she invests effort. Every worker can have a different proficiency matrix.

\begin{definition}[Trustworthiness Matrix]The trustworthiness matrix $T_i$ of worker $i$ is a $K \times K$ right stochastic matrix, where $T_i[g,y]$ ($\forall g,y \in[K]$) is probability of her reported answer on a task being $y$ given that ground truth is $g$.
\end{definition}

Note the difference between the proficiency and trustworthiness matrices. Proficiency models worker's ability while trustworthiness is a function of her ability and honesty.

\begin{proposition}\label{prop:basic}
	If $e_i = 1$, the trustworthiness matrix $T_i$ of a worker $i$ is given by $T_i = A_i S_i$. If $e_i = 0$, $T_i$ is a matrix with all rows equal to reporting strategy vector $\vec{S_i}$.
\end{proposition}

\begin{figure}[ht]
	\centering
	\includegraphics[width=0.48\textwidth]{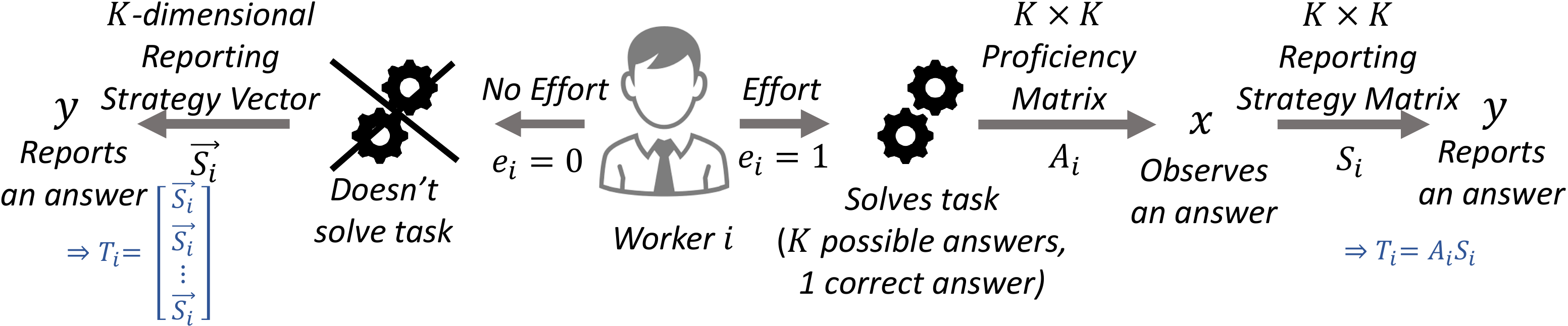}
	\caption{Model}
	\label{fig:model}
\end{figure}

Our model is summarized in Figure~\ref{fig:model}.

Until now, we defined the strategy space for the settings in which workers solve one task each. In our work, we consider settings, in which workers solve multiple tasks and next define the strategy space for multi-task settings.
\begin{definition}[Uniform Strategies for Multi-task Settings]
	A worker's strategy (effort and reporting) in multi-task settings is called uniform if the strategy is the same on all tasks in a given batch solved by the worker.
\end{definition}

The uniform strategies are sometimes also called consistent strategies in the literature. It is important to note that the space of uniform strategies DOES INCLUDE mixed strategies. The motivation for considering only uniform strategies in the multi-task literature is that the tasks of similar nature can be grouped in batches so that workers don't strategically distinguish between tasks assigned to them.

\begin{definition}[Dominant Uniform Strategy Incentive Compatibility]
	Given that workers can play any strategy from the space of uniform strategies, an incentive mechanism is called dominant uniform strategy incentive compatible (DUSIC) if the expected reward of any worker is strictly maximized by playing a truthful strategy, no matter what uniform strategies other workers use.
\end{definition}

In this notion of incentive compatibility, the reward of a worker is \textit{strictly} maximized in a truthful strategy even if others are not truthful. Thus, the truthful strategy dominates any heuristic strategy of not solving the tasks and non-truthful strategies of solving the tasks but reporting non-truthfully. It also dominates any mixed uniform strategy.

\begin{definition}[Oracle]
An agent $o$ is called an oracle if her trustworthiness matrix $T_o$ is known and $T_o$ doesn't have identical rows.
\end{definition}
For example, if the oracle is the source of gold standard answers, then by definition of gold tasks, oracle's trustworthiness matrix is an identity matrix.

We use $P(g)$ to denote the prior probability of the ground truth answer of any randomly selected task being $g$. It is assumed to be known and fully mixed ($P(g) > 0\ \forall g \in [K]$). It can also be estimated from the gold standard answers.

\section{Finding Trustworthiness Transitively}\label{sec:prelims}
In this section, we first explain the main building block of our mechanism: the process of finding the trustworthiness of a worker, given the trustworthiness of another worker by using the joint distribution of their answers on shared tasks.
\begin{definition}[Peer]\label{def:peer}
	For a worker $i$, the mechanism assigns another worker $j$ as her peer. Workers $i$ and $j$ are assigned sets of tasks $Q^i$ and $Q^j$ respectively such that $|Q^i \cap Q^j| \gg 0$.
\end{definition}
The definition requires that some tasks are solved by both the worker and her peer. Both workers also solve some other tasks that are not shared. It may be noted that eliciting answers of multiple workers on same tasks is the central idea in crowdsourcing~\cite{surowiecki2005wisdom} and is not a new requirement introduced in our paper.

Let $T_j$ be the known trustworthiness matrix of worker $j$ and let $j$ be the peer of another worker $i$, whose trustworthiness matrix $T_i$ is not known. We want to find the unknown $T_i$ using the answers given by the two workers and the known $T_j$. Since the worker $i$ and her peer $j$ solve some shared tasks by definition, their reported answers on these shared tasks provide the mechanism with an empirical joint distribution of their answers. We use $\omega(Y_i = y_i|Y_j = y_j)$ to denote this conditional empirical distribution and $\omega(Y_j = y_i)$ to denote the empirical distribution of answers of peer $j$ only.

\begin{lemma}~\label{lemma:general}
	As $|Q^i \cap Q^j| \rightarrow \infty$, the following holds w.h.p.
	\begin{equation}~\label{eq:lemma-general}
		\omega(Y_i = y_i | Y_j = y_j) = \sum_{g \in [K]} T_i[g, y_i] \cdot \Big(\frac{T_j[g, y_j]\cdot P(g)}{\omega(Y_j = y_j)}\Big)
	\end{equation}
	$\forall\ y_i, y_j \in [K]$ and  $\ \omega(Y_j = y_j) \neq 0$.
\end{lemma}
The proof of Lemma~\ref{lemma:general} is provided in the supplementary material. The LHS in Equation~\ref{eq:lemma-general} is the conditional probability $P(Y_i = y_i | Y_j = y_j)$ in the limit. When we apply Bayes' rule to write this conditional probability in terms of other model probabilities, we get the RHS of Equation~\ref{eq:lemma-general}. This assumes that the answers of workers $i$ and $j$ are conditionally independent given the ground truth.

In the linear system of Equations~\ref{eq:lemma-general}, $T_i[g, y_i]$ $\forall\ g, y_i \in [K]$ are unknowns. Since the matrix $T_i$ is also right stochastic, we have as many equations as the number of unknowns. This system can be solved for $T_i$, provided the system is \textit{well-defined}. This requires that $\omega(Y_j = y_j) \neq 0$ and for a unique solution, the coefficient matrix of this linear system must have linearly independent rows. This system of linear equations can be solved analytically. In practice, many libraries are also available for computing the solution efficiently. We now use this transitive method of finding unknown $T_i$ to develop our mechanism in the next section.
\begin{figure*}[ht]
	\begin{minipage}{1\textwidth}
		\rule{1\textwidth}{0.4pt}
		\textbf{Mechanism 1 : The Deep Bayesian Trust Mechanism} \\
		\rule{1\textwidth}{0.4pt}
		\begin{enumerate}[leftmargin=*]
			\item \label{step1-b} Assign a set of tasks to the oracle $o$ and obtain its answers on the tasks.
			\item \label{step2-b} Initialize an \textit{Informative Answer Pool} (IAP) with the answers given by oracle. \\	\noindent\fcolorbox{light-gray}{light-gray}{%
				\begin{minipage}{0.964\textwidth}
					$IAP = \Big[ \big[o : T_o : q_1-\mathbf{a_1}, q_2-\mathbf{a_2}, q_3-\mathbf{a_3}, ... \big] \Big]$
					
					$o$ stands for oracle, $T_o$ is the trustworthiness of the oracle and $q_l-\mathbf{a_l}$ are the task $-$ answer pairs provided by the oracle.
			\end{minipage}}%
			
			\item\label{step3-a} Select some tasks submitted by a worker from the IAP. If there is no worker yet in the IAP, select the oracle's tasks.
			
			\item\label{step3-b} Prepare a set of batches of tasks such that each batch contains tasks selected in the previous step. Mix some fresh tasks in each of the batch.
			
			\item\label{step3-c} Publish the batches on the platform and let workers self-select themselves to solve one batch each.
			
			\item\label{step4-b} For any worker $i$ who submits her batch, solve the system of Equations in~\ref{eq:lemma-general} to find the unknown trustworthiness $T_i$. Reward worker $i$ for her answers with an amount equal to $\beta\cdot R_i$ where, $R_i =  \Big(\sum\limits_{g \in [K]}T_i[g,g]\Big)  - 1$.
			\item \label{step5-b} If the answers of worker $i$ satisfy the informativeness criterion, add the answers to the IAP and assign them trustworthiness $T_i$ as obtained in Step~\ref{step4-b}.\\
			\noindent\fcolorbox{light-gray}{light-gray}{%
				\begin{minipage}{0.964\textwidth}
					For example, at a given instant, the pool may look as follows : 
					
					$AP = \Big[ \big[o :  T_o : q_1-\mathbf{a_1}, q_2-\mathbf{a_2}, ... \big], \big[W_1 : T_{W_1} : q_2-\mathbf{a_2}, q_4-\mathbf{a_4}, ....\big], \big[W_2 : T_{W_2} : q_2-\mathbf{a_2}, q_5 -\mathbf{a_5}, ...\big], ... \Big]$ 
					
					Here, $W_i$ are the identities of workers followed by their trustworthiness $T_{W_i}$ and their submitted task-answer pairs.
			\end{minipage}}%
			\item \label{step6-b} \textit{Asynchronously} repeat steps~\ref{step3-a},~\ref{step3-b},~\ref{step3-c},~\ref{step4-b} and \ref{step5-b} whenever any worker submits her batch, until desired number of answers are collected for all tasks.
		\end{enumerate}
		\rule{1\textwidth}{0.4pt}
	\end{minipage}
	\label{fig:mechanism}
\end{figure*}

\section{The Deep Bayesian Trust Mechanism}
The Deep Bayesian Trust mechanism is summarized in Mechanism 1 on the next page. It maintains a pool of workers' answers which are ``informative" for evaluating other workers. The meaning of the term\textit{ informative} will be explained later. The pool is initialized with some tasks and their answers given by the oracle. In crowdsourcing terminology, these are the gold task-answer pairs. The trustworthiness matrix of the oracle is initialized to be $T_o$.  Since by definition, gold tasks are the tasks whose \textit{correct} answers are known, $T_o$ is an identity matrix. The mechanism then publishes several batches of tasks on the platform such that each batch has some tasks in common with the tasks solved by the oracle and some unique new tasks in each batch. Workers self-select themselves to solve one batch each and report their answers for respective batches. Thus, the oracle becomes the peer of each of these workers. 

\noindent\fcolorbox{light-gray}{light-gray}{\begin{minipage}{0.46\textwidth}Let's assume that the oracle solves $s_o$ number of tasks. The mechanism publishes $k$ batches of tasks such that there are $s_o$ tasks in common with the oracle and $s_n$ unique new tasks in each batch. Thus, it publishes $k \cdot s_n$ tasks that are new (not solved by the oracle already) and also $k$ instances of the same $s_o$ tasks that are already solved by the oracle. $k$ becomes a hyper-parameter of the mechanism and $s_o + s_n$ becomes the size of the batches solved by every worker.\end{minipage}}
As the workers start submitting their respective batches, the mechanism also starts rewarding the workers for their answers, asynchronously (without waiting for other workers). To calculate the reward, the mechanism uses Lemma~\ref{eq:lemma-general} for finding the trustworthiness matrix of the answers given by workers. Note that the lemma is applicable because the trustworthiness of the peer (oracle) is known. The reward for worker $i$ is given by $\beta\cdot R_i$, where $R_i =  \Big(\sum\limits_{g \in [K]}T_i[g,g]\Big)  - 1$ and $\beta$ is a scaling constant. 
\noindent\fcolorbox{light-gray}{light-gray}{\begin{minipage}{0.46\textwidth} $R_i$ takes the summation of the diagonal entries of the trustworthiness matrix $T_i$. These are the accuracy parameters of the worker (the probabilities of the workers' answers being same as the ground truth). $R_i$ further subtracts $1$ from this summation for technical reasons that will be exploited later to ensure a desired incentive property.
\end{minipage}}
The worker gets her reward and is out of the mechanism. At this stage, the mechanism decides whether to reuse the answers given by the worker for evaluating more workers. If the worker's answers satisfy a certain ``informativeness" criterion, they are added to the pool. If the worker's answers are added to the pool, the mechanism can immediately publish more batches such that there are some tasks in common with the new (non-gold) tasks just solved by the previous worker and some more new tasks in each of the batches. This step is the same as described earlier. The only difference is that now the batches being published have tasks in common with the tasks solved by a worker, not the oracle (i.e. the peers are now other workers, not the oracle). These steps are repeated in parallel and asynchronously until the mechanism has obtained the desired number of answers for all its unsolved tasks.

\begin{figure}[ht]
	\centering
	\begin{subfigure}{0.05\textwidth}
		\centering
		\includegraphics[width=1\textwidth, page = 1]{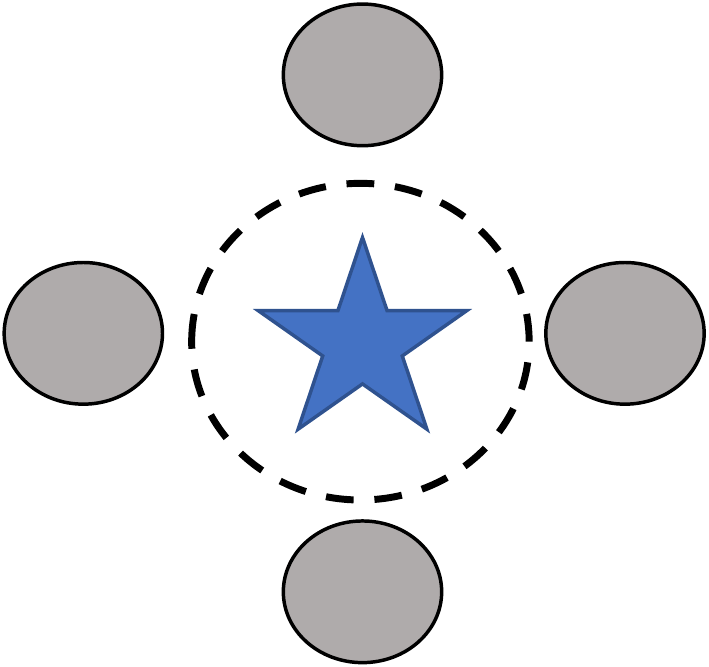}
	\end{subfigure}
	\dashline
	\begin{subfigure}{0.05\textwidth}
		\centering
		\includegraphics[width=1\textwidth, page = 2]{DBT-cropped.pdf}
	\end{subfigure}
	\dashline
	\begin{subfigure}{0.09\textwidth}
		\centering
		\includegraphics[width=1\textwidth, page = 3]{DBT-cropped.pdf}
	\end{subfigure}
	\dashline
	\begin{subfigure}{0.09\textwidth}
		\centering
		\includegraphics[width=1\textwidth, page = 4]{DBT-cropped.pdf}
	\end{subfigure}
	\begin{subfigure}{0.14\textwidth}
		\centering
		\includegraphics[width=1\textwidth, page = 5]{DBT-cropped.pdf}
	\end{subfigure}
	\caption{Illustration of the Deep Bayesian Trust Mechanism}
	\label{fig:mech_pict}
\end{figure}

To summarize (Figure~\ref{fig:mech_pict}), the mechanism starts with an answer pool seeded with the oracle's answers, uses the answers in the pool to assess trust in other workers' answers, expands this pool based on the informativeness of the workers' answers and repeats the process. We emphasize that the mechanism doesn't assign any permanent reputation to workers. A worker's answers being added to the pool is not the same thing as a worker being ``pre-screened" and certified trusted. We just evaluate the answers provided by a worker in any given batch and add them to the answer pool together with an estimate of the trustworthiness of that batch of answers.

\subsection*{Informativeness Criterion}
We can now discuss the informativeness criterion for workers' answers, which was omitted earlier. The purpose of the informativeness criterion is to check whether the answers provided by a worker $i$ can be used to estimate the trustworthiness of another worker or not. As discussed in Section~\ref{sec:prelims}, this depends on whether the coefficient matrix of the linear system of equations has linearly independent rows or not. For example, assume that the answers of worker $j$ are added to the pool and let $i$ be another worker who gets $j$ as her peer in the future. In that case, the mechanism will solve the following equations to estimate the trustworthiness of $i$: 

\begin{small}$$\omega(Y_i = y_i | Y_j = y_j) = \\ \sum_{g \in [K]} T_i[g, y_i] \cdot \Big(\frac{T_j[g, y_j]\cdot P(g)}{\omega(Y_j = y_j)}\Big)$$
$\forall\ y_i, y_j \in [K]$\end{small}\\

The coefficients $\frac{T_j[g, y_j]\cdot P(g)}{\omega(Y_j = y_j)}$ of this linear system don't depend on the answers given by worker $i$ and the mechanism can determine in advance whether the system will be solvable by just looking at these coefficients. If $\omega(Y_j = y_j) \neq 0$ and the coefficient matrix is full rank, the informativeness criterion is said to be satisfied and the answers of worker $j$ are added to the pool. It may be noted that the informativeness criterion doesn't require answers of only truthful or high proficiency workers to be added to the pool. Answers from non-truthful or low proficiency workers can also be added to the pool.

\noindent\fcolorbox{light-gray}{light-gray}{\begin{minipage}{0.46\textwidth} To understand this technical criterion in more depth, note that the coefficients $\frac{T_j[g, y_j]\cdot P(g)}{\omega(Y_j = y_j)}$ of the linear system are equal to the posterior distribution $P(G = g | Y_j = y_j)$ by Bayes' rule. Thus, the answers of a worker $j$ satisfy the informative criterion if the posterior distributions $P(G = g | Y_j = y_j)$ over $g \in [K]$ for any two different $y_j \in [K]$ are not identical. One interesting example, where the informative criterion is not satisfied, is when the peer $j$ plays a heuristic strategy. In such a case, the reported answers are not correlated with the ground truth and the posterior distributions $P(g | Y_j = y_j)$ are same as the prior distribution $P(g)$ or in other words, the reported answers are not ``informative" of the ground truth. 
		
It may be noted that the informativeness criterion is fairly weak for the binary answer spaces. It only requires that the reports of the peer $j$ are not independent of the ground truth ($T_j[1,1] + T_j[2,2] \neq 1 \implies T_j[1,1] \neq T_j[2,1] \implies T_j[2,2] \neq T_j[1,2]$).\end{minipage}}

\section{Analysis}
We now prove strong game-theoretical properties for our mechanism. In this discussion, we will assume that a worker and her peer solve many shared tasks ($|Q^i \cap Q^j| \rightarrow \infty$). This is \textbf{not} the same as requiring every task to be solved by large number of workers, which would have been inefficient. In later sections, we will also discuss the empirical performance of our mechanism without this assumption. We use $C^E$ to denote the cost of effort required to solve a batch of tasks. Proofs are provided in the supplementary material.

\begin{theorem}\label{thm:main}
	If $\beta > \frac{C^E}{\big(\sum\limits_{g \in [K]}A_i[g,g]\big) - 1}$ and $A_i[g,g] > A_i[g^\prime, g],\ \forall g^\prime \neq g$, then the Deep Bayesian Trust mechanism
	
	\begin{enumerate}[leftmargin=*, label=(\roman*)]
		\item is dominant uniform strategy incentive compatible (DUSIC) for every worker $i$; 
		
		\item ensures strictly positive expected reward in the truthful strategy.
	\end{enumerate}
\end{theorem}
Theorem~\ref{thm:main} requires a condition on the scaling constant $\beta$ to cover the cost of effort, and reduces to $\beta > 0$ when cost of effort is $0$. The condition required on proficiency matrix $A_i$ (i.e. $A_i[g,g] > A_i[g^\prime, g],\ \forall g^\prime \neq g$)\footnote{For binary answer space, the theorem can also be shown to hold under a weaker condition $A_i[0,0]+A_i[1,1] > 1$.} can be more easily understood in the case of binary answers. In binary settings, the condition is satisfied if $A_i[0,0] > 0.5$ and $A_i[1,1] > 0.5$. This is \textbf{not} a condition on the honesty of the workers but only on their ability. The condition merely ensures that the worker can obtain answers that are positively correlated with the ground truth. Such conditions are common in the literature~\cite{dasgupta2013crowdsourced}. Unlike the literature, the condition here only affects the best strategy of a given worker, not of all the workers. For example, if the condition is not satisfied for a worker, she may find it better to deviate to a non-truthful strategy but it doesn't affect the dominant strategy of other workers. We note that such informed deviation by a low proficiency worker to increase the accuracy of her reported answers is not bad for the requester.

\begin{corollary}\label{coro:scaling}
 The scaling constant $\beta$ of the Deep Bayesian Trust mechanism is independent of the probability of a worker getting oracle or another truthful worker as peer.
\end{corollary}
Corollary~\ref{coro:scaling} implies that to ensure incentive compatibility, our mechanism doesn't need to scale up the rewards of workers if the probability of a worker getting oracle or another truthful worker as peer decreases.

\begin{theorem}\label{thm:heur}
	In the Deep Bayesian Trust mechanism, a heuristic strategy gives zero expected reward.
\end{theorem}
It may be noted that the DUSIC result in Theorem~\ref{thm:main} already implies that the heuristic strategies are not in equilibrium but Theorem~\ref{thm:heur} answers the question that what if someone still plays those strategies.

\subsubsection{Limitation} If, despite all these guarantees, every single worker chooses to \textit{irrationally} play a heuristic strategy, then our mechanism will not be able to expand its pool and will be forced to behave like other mechanisms which assign gold tasks to every worker. But (i) such workers don't gain anything from the mechanism; (ii) the dominant incentive compatibility of the mechanism remains unaffected for any \textit{rational} workers even in such a degenerate case; and (iii) the heuristic strategy doesn't become an equilibrium strategy.

\subsection{Fairness of Rewards} Recently, concerns have been raised about fairness and other ethical considerations in algorithms that affect humans~\cite{acm-statement,president-statement2}. The discussion on fair rewards in crowdsourcing has included issues such as minimum wages and adequate compensation for time and effort~\cite{schmidt2013good} but there has not been any principled approach to address the issue of fairness in rewards from a non-discrimination perspective. For example, if a worker with higher ability gets a lower reward than a worker with lower ability because of the difference in the way they were evaluated, then this is a potential case of unfairness. The unfairness is an unintentional and undesired property of the existing mechanisms. Peer based mechanisms in the literature randomly select peers and reward the workers based on their answers and the answers of their respective peers. The reward of the workers is generally a function of their own ability as well as their peers' ability, making the rewards unfair. This unfairness issue in the peer based mechanisms was first pointed out by~\cite{kamar2012incentives}.
The issue becomes more serious when workers know ex-ante that they are being evaluated using peers with different proficiencies. This is the case, for example, with the mechanism of~\cite{de2016incentives}. Our mechanism doesn't need to inform the workers about their peers at all but as we show, the mechanism can satisfy an even stronger definition of fairness.
\begin{definition}[Fair Incentive Mechanism]
	An incentive mechanism is called fair if the expected reward of any worker is directly proportional to the accuracy of the answers reported by her and independent of the strategy and proficiency of her random peer.
\end{definition}
This is a reasonable definition of fairness and is in agreement with the broader theory for \textbf{\textit{individual fairness}} of algorithms. For example, the pioneering work of~\cite{dwork2012fairness} defines that fair algorithms take similar decisions for individuals with similar relevant attributes. The relevant attribute in our case is the worker's accuracy. The definition is also non-trivial to satisfy. In existing peer based mechanisms, the rewards also depend on the unknown ability of the peer (even if the peer can be believed to be truthful). For example, in the mechanism of~\cite{dasgupta2013crowdsourced}, the reward of a worker in the truthful equilibrium is an increasing function of her proficiency as well as her peer's proficiency. On the contrary, our mechanism satisfies this definition of fairness. The mechanism carefully uses the peer answers only to find trustworthiness of a worker, which is completely her own accuracy parameter and doesn't depend on her peer's proficiency or strategy.

\begin{theorem}
	The Deep Bayesian Trust Mechanism is fair.
\end{theorem}
This is perhaps a surprising result because in the existing framework of the peer based mechanisms, one would perhaps reason that it is impossible for the rewards to not depend on the accuracy of the peer.

\section{Numerical Simulations}
In this section, we evaluate the performance of our mechanism empirically. We simulate the settings in which workers with different proficiencies $A_i$ report answers to different tasks. The proficiency matrices of different workers were generated independently such that the diagonal entries $A_i[g, g]\ \forall g \in [K]$ were $\beta(5,1)$ distributed. The diagonal entries $A_i[g, g]\ \forall g \in [K]$ for a given worker $i$ are not necessarily the same as they are also independently generated. Rest of the entries are generated randomly such that every row of proficiency matrix sums to $1$.

We consider following strategies that workers may play:
\begin{enumerate}[leftmargin=*]
	\item \textbf{Truthful}- Workers obtain answer for any given task based on their respective proficiency matrices and report the answers truthfully.
	\item \textbf{Heuristic} - Workers' reported answers are generated independently of their proficiency using a common distribution over the answer space.
	\item \textbf{Permutation} - Workers obtain answer for any given task based on their respective proficiency matrices but they apply a common permutation on the answers before reporting it to the mechanism. In a non-truthful permutation deterministic strategy~\cite{Shnayder2016}, workers solve the tasks, but they apply a permutation mapping on the answers before reporting it to the mechanism. For example, in a ternary answer space ($K = 3$), a permutation $f$ can be as follows :
	$f(0) = 1, f(1) = 2, f(2) = 0$, i.e., whenever the obtained answer is $0$, workers report $1$, for $1$, they report $2$ and for $2$, they report $0$. In a binary answer space, this corresponds to reporting the opposite of the obtained answer.
\end{enumerate}

In general, the simulations performed in the literature for peer based mechanisms compare the average reward in different equilibria. For example, the average reward of workers when all of them play a truthful strategy may be compared with the average reward when all play a heuristic strategy. This is because such mechanisms only guarantee that different strategies are in equilibria and that one equilibrium is more profitable than the other. But our stronger theoretical result (dominant incentive compatibility) demands stronger simulations. We go beyond comparing just equilibrium rewards and instead compare the rewards of workers playing different strategies against one another at the same time. More precisely, in our simulations, we don't require every worker to play a common strategy. Any worker can play a heuristic, permutation or truthful strategy with equal probability. Such settings can't be handled by mechanisms that guarantee only equilibrium results. We will show that in our mechanism, there is a clear distinction between the rewards of workers playing different strategies with truthful workers being nicely rewarded and others being penalized. Workers were simulated to be hired in $4$ rounds, with $5$, $25$, $125$ and $625$ workers in successive rounds. $K$ was set to $2$ in all the experiments discussed in the paper. \begin{figure}[ht]
	\centering
	\includegraphics[width=0.45\textwidth]{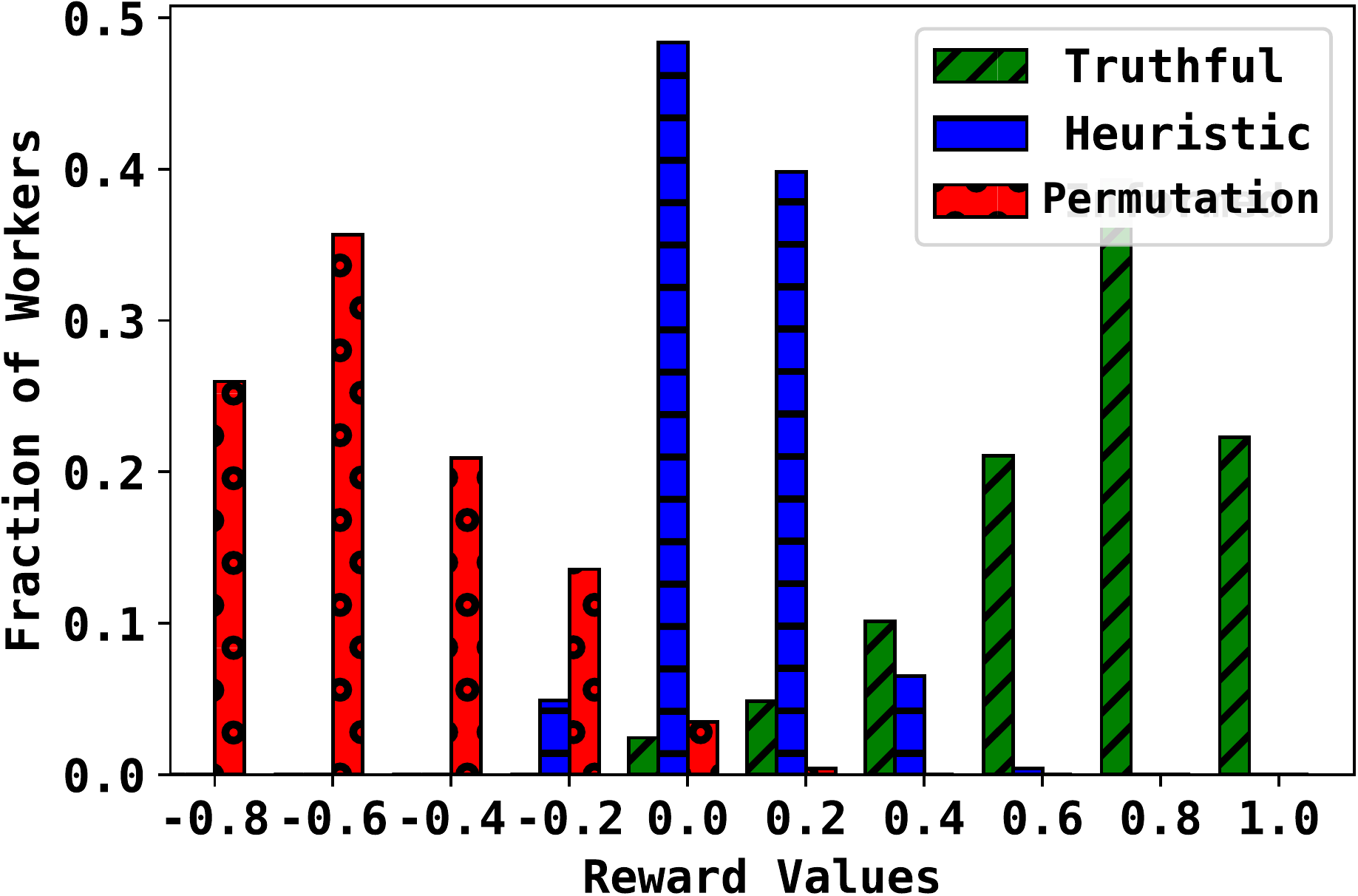}
	\caption{Distribution of rewards for workers with different proficiencies playing different strategies}
	\label{fig:beta-rewards}
\end{figure}

Figure~\ref{fig:beta-rewards} compares the distribution of rewards of workers playing the three strategies. The rewards of the workers playing the heuristic strategy are centered around $0$, as expected from Theorem~\ref{thm:heur}. The reward of workers playing truthful strategy are centered around a strictly positive value as predicted by Theorem~\ref{thm:main}. On the contrary, the rewards of workers playing the permutation strategy are symmetrically centered around a strictly negative value. It may be noted that in existing peer based mechanisms, permutation strategies (in equilibria) are equally profitable as the truthful strategy, which is clearly not the case with our mechanism. Firstly neither heuristic nor permutation strategies are in equilibrium in our case and even if workers use any of these strategies, they get lower reward than the truthful strategy.
\begin{figure}[ht]
		\centering
		\includegraphics[width=0.45\textwidth]{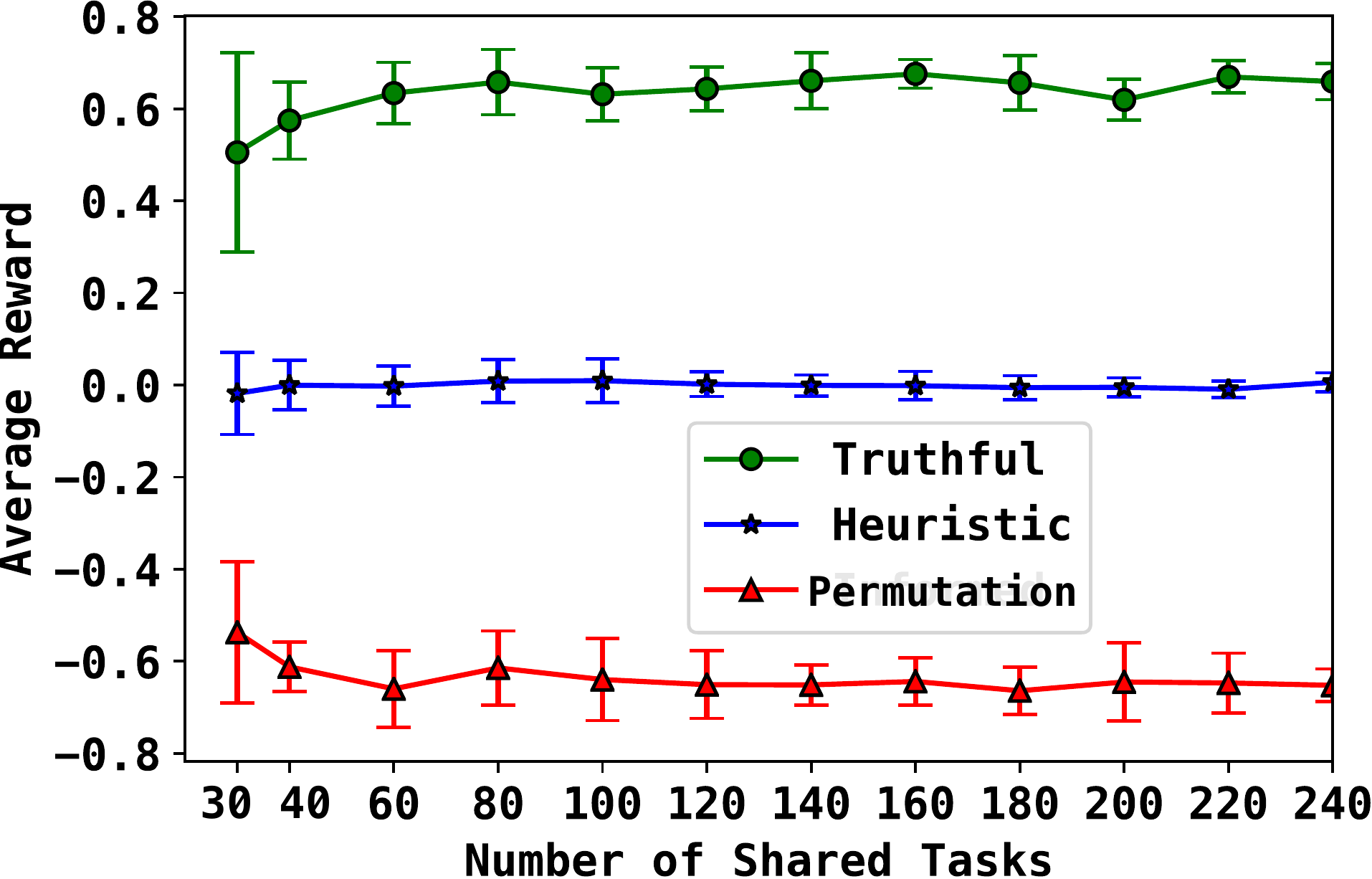}
		\caption{Average reward of workers playing different strategies under different number of shared tasks}
		\label{fig:beta-robust-rewards}
\end{figure}

We now show the \textbf{robustness of our mechanism with respect to the number of shared tasks between workers}. We discussed only the asymptotic properties of the mechanism earlier in the theoretical analysis. Hence, this simulation study is important to show the performance of the mechanism with a finite number of shared tasks. Figure~\ref{fig:beta-robust-rewards} compares the average of rewards of the workers (with $\beta(5,1)$ distributed proficiencies) playing different strategies under different settings of the number of shared tasks. Error bars show the standard deviation in $100$ repeated runs. The trend discussed in previous experiment can be observed to be very robust to the number of shared tasks. Thus, the Deep Bayesian Trust mechanism is attractive even when the number of shared tasks is not large. This simulation also implies that with only $30$ gold tasks (and given only to $5$ workers), the mechanism can reward $5+25+125+625+...$ workers. 

We also simulated the settings in which the diagonal entries $A_i[g, g]\ \forall g \in [K]$ were uniformly distributed in $(\frac{1}{K},1]$ and repeated the above experiments. Results (with similar observations) are available in the supplementary material. 

We also conducted a preliminary study on Amazon Mechanical Turk to observe the effect of our mechanism in encouraging workers to invest more effort. Workers were given hard tasks related to natural language understanding, and they had to given a binary (`Yes' or `No') answer. The details of the study are in the supplementary material. We observed that in presence of our mechanism, the workers invested more time in solving the tasks and the average accuracy of their responses also improved.

\section{Conclusions}
We proposed the Deep Bayesian Trust mechanism to incentivize crowdworkers in large scale settings. The mechanism rewards the workers for the correctness of their reports without checking every worker with gold tasks. Instead, it uses the correlation in the answers of the workers and their peers to estimate their accuracy. The mechanism is guaranteed to be game theoretically robust to any strategic manipulation. Thus, it is also suitable even for the settings in which workers of very heterogeneous proficiencies and motivations solve the tasks at the same time. The mechanism also ensures fair rewards to workers, thus contributing towards the bigger movement of making algorithmic decisions fair. Among other issues, our mechanism notably addresses the scalability issues in purely gold tasks based mechanisms, the incentive compatibility issues in purely peer based mechanisms and the information requirement and fairness issues in the mixed mechanisms.
\bibliographystyle{aaai}
\bibliography{aaai19}
\end{document}